\documentclass[11pt, amssymb, amsmath, showpacs, aps,  cha]{revtex4-1} 
\usepackage{graphicx,color,indentfirst,amsthm,ulem,bm,bbold,subfig,caption,amsmath,float}
\usepackage{natbib}
\begin{document}
\title{Collective resonance light scattering from thermally relaxing systems in cavities}

\author{Bingyu Cui$^{1}$}
\email{bycui@cuhk.edu.cn}
\affiliation{${}^1$School of Science and Engineering, The Chinese University of Hong Kong, Shenzhen, Guangdong, 518172, P. R. China}
\date{\today}

\begin{abstract}
\noindent We study steady-state resonance light scattering from ensembles of noninteracting molecules, both in free space and inside optical cavities, while accounting for local thermal relaxation. The scattering spectra are obtained from steady-state solutions of either the Schrödinger equation or a Liouville-space master equation. In the absence of a cavity, the spectra exhibit an elastic peak at the incident-photon energy and an inelastic fluorescence peak near the molecular excitation energy. Inside a cavity, the fluorescence peak splits into upper- and lower-polaritonic peaks in the strong-coupling regime. We analyze how the elastic and inelastic spectral features scale with the number of molecules under fixed cavity-molecule coupling and identify distinct collective trends in the Rayleigh peak intensity and in the integrated polaritonic or fluorescence spectral weight. The two theoretical approaches yield qualitatively consistent results while highlighting different aspects of thermally induced relaxation and dephasing.
\end{abstract}
\maketitle

\section{Introduction}
Light-matter interactions provide some of the most powerful probes of molecular structure and dynamics. In particular, optical spectroscopy reveals how matter absorbs, emits, and scatters electromagnetic radiation, thereby providing information about molecular energy levels, relaxation pathways, and environmental effects \cite{Pavia2014}. When weak monochromatic light interacts with a molecular system, several outcomes are possible: the light may be transmitted, absorbed and later re-emitted as fluorescence, scattered elastically, or scattered inelastically \cite{Gardiner1989}. Elastic scattering preserves the photon energy and includes Rayleigh scattering, whereas inelastic scattering changes the photon energy through exchange with internal molecular degrees of freedom. In molecular spectroscopy, fluorescence and Raman scattering are especially important because they probe different aspects of excitation and relaxation dynamics.

Fluorescence is an incoherent emission process in which the molecule first absorbs a photon and is promoted to an excited electronic state, then relaxes and emits at a later time. Because the excited molecule typically undergoes rapid internal relaxation before radiative decay, fluorescence often occurs predominantly from the lowest excited electronic state, consistent with Kasha’s rule \cite{Kasha1950}. As a result, fluorescence spectra are usually much less sensitive to the excitation frequency than coherent scattering signals. By contrast, coherent elastic and Raman scattering retain a direct relation to the driving field and therefore encode different information about the excitation pathway.

In recent years, the optical response of molecules in structured electromagnetic environments has attracted great attention. When molecular systems are placed near dielectric interfaces, plasmonic nanostructures, or optical cavities, the local electromagnetic field and the molecular response can be strongly modified \cite{Kristensen2013,Chikkaraddy2016}. In the strong-coupling regime, hybrid light-matter states—polaritons—emerge, and their formation can substantially alter optical spectra and relaxation dynamics. This general framework has motivated extensive work in molecular polaritonics and cavity quantum electrodynamics, with proposed or demonstrated consequences for energy transport \cite{Feist2015,Zhong2016,Zhong2017,Rozenman2017,Xiang2020,Georgiou2021,Schachenmayer2015,Schfer2019,Aeschlimann2017,Du2018,Rustomji2019}, charge transport and transfer \cite{Orgiu2015,Krainova2020,Halbhuber2020,David2017,Hagenmller2018,Shan2019,Pang2020,Mandal2020,Mauro2021}, heat transport \cite{Yang2020}, electronic structure \cite{Flick2018}, molecular potential-energy surfaces \cite{Flick2017,Christian2018,Flick2019,Haugland2020,Haugland2021,Luk2017}, and chemical reactivity \cite{Hoffmann2020,Schfer2022,Hutchison2012,Galego2016,Munkhbat2018,Mandal2019,Herrera2016}.

Of particular interest are collective effects in molecular ensembles strongly coupled to one or a few cavity modes. When many molecules couple coherently to the same confined radiation field, the optical response may depend strongly on the number of participating molecules. Classic examples include Dicke superradiance \cite{Dickle1954} and the collective Rabi splitting of upper and lower polaritons, whose magnitude scales as $\sqrt{N}$ for $N$ identical emitters coupled to a single cavity mode \cite{Li2022,Cheng2023}. These collective features have stimulated broad interest, ranging from polariton-enabled transport and chemistry to proposals for collective energy storage and quantum batteries \cite{Alicki2013,Ferraro2018}. At the same time, the interpretation of experimentally observed collective effects in molecular spectroscopy remains subtle, especially in the presence of disorder, thermal relaxation, and dephasing \cite{Dunkelberger2022,Ying2026}. In particular, it is still important to clarify which spectral signatures reflect genuinely collective radiative behavior and which arise from local relaxation processes that remain essentially single-molecule in character.

This issue motivates the present work. We investigate long-time, energy-resolved scattering spectra of noninteracting molecular ensembles both in free space and in optical cavities. Our model contains molecular excited states coupled to incident and outgoing radiation channels, with additional environmental effects incorporated in two complementary ways. In the first approach, thermal fluctuations are represented by stochastic modulation of the molecular excitation energies. In the second, relaxation and dephasing are treated phenomenologically within a density-matrix description. Although these two formulations differ in their microscopic assumptions, they lead to qualitatively consistent steady-state spectra and therefore provide a useful basis for comparing collective signatures. These approaches allow us to identify how elastic and inelastic scattering features scale with the number of molecules and how these trends are modified by cavity coupling. Outside the cavity, we identify an elastic peak at the incident energy and an inelastic fluorescence peak near the molecular excitation energy. Inside the cavity, the fluorescence feature splits into upper- and lower-polaritonic peaks in the strong-coupling regime. We then analyze how the heights and integrated spectral weights of these features vary with ensemble size, thereby distinguishing collective radiative enhancement from incoherent thermally induced response.

The remainder of this paper is organized as follows: Section II introduces the model and the two theoretical approaches used to calculate the scattering spectra. Section III presents the numerical results. Section IV discusses the physical interpretation of the observed scaling behaviors and the differences between the two methods. Section V summarizes the main conclusions.

\section{Methods}

Information about light scattering can be obtained under continuous-wave (CW) excitation, where a long-duration incident field with well-defined frequency and propagation direction drives the system. The scattered light is then monitored in the long-time limit as a function of outgoing energy and direction. Below, we consider a model consisting of a collection of noninteracting molecular excited states coupled to an incident radiation mode and to a continuum of outgoing photon states.

We denote the molecular excited states by $|k\rangle$, with excitation energies $E_k$ for $k=1,...,N$. The incident field is represented by a pumping state $|in\rangle$, and the outgoing radiation is described by a continuum of states $|j\rangle$ \footnote{In fact, the state $|in\rangle$ formally belongs to the group of radiative continua, but it (and the state $|out\rangle$ denoted later) has a special status because it represents the incident (outgoing) states of the process under discussion. An alternative interpretation of $|in\rangle$ and $|j\rangle$ is that they refer to the initial state of unexcited molecules dressed by the incident field at a particular state and final state of unexcited molecules with the outgoing photons in another state.}. The Hamiltonian is
\begin{align}
    \hat{H}_{SS}&=E_{in}|in\rangle\langle in|+\sum_{k=1}^N\left[E_k|k\rangle\langle k|+V_{in,k}|in\rangle\langle k|+V_{k,in}|k\rangle\langle in|\right]\notag\\
    &+\sum_{j}\left[E_j|j\rangle\langle j|+\sum_{k=1}^N\left(V_{k,j}|k\rangle\langle j|+V_{j,k}|j\rangle\langle k|\right)\right],
    \label{eq:HSSF}
\end{align}
Here we use the rotating-wave approximation and take $V_{in,k}=V_{k,in}^*,V_{k,j}=V_{j,k}^*$. These couplings represent the interaction of the molecular excitation with the incident and outgoing radiation modes through the corresponding dipole matrix elements.

The wavefunction may be expanded as
\begin{equation}
    \Psi(t)=c_{in}(t)|in\rangle+\sum_{k=1}^Nc_k(t)|k\rangle+\sum_jd_j(t)|j\rangle,
\end{equation}
and the Schr\"{o}dinger equation  $i\hbar\dot{\Psi}=\hat{H}_{SS}\Psi$, leads to the amplitude equations
\begin{align}
    i\hbar\frac{d}{dt}\left(\begin{matrix}
        c_{in}\\
        ...\\
        c_k\\
        ...\\
        d_j
    \end{matrix}\right)=\left(\begin{matrix}
        E_{in}c_{in}+\sum_kV_{in,k}c_k\\
        ...\\
        E_kc_k+V_{in,k}c_{in}+\sum_jV_{k,j}d_j\\
        ...\\
        E_jd_j+\sum_kV_{k,j}c_k
    \end{matrix}\right).
    \label{eq:schSSF}
\end{align}

We are interested in the weak-driving, long-time steady state. To represent CW excitation, we impose the driving boundary condition $c_{in}=\tilde{c}_{in}\exp(-iE_{in}t/\hbar)$ with constant amplitude $\tilde{c}_{in}$. In addition, each outgoing continuum state is assigned a small damping rate $\eta/2$, which enforces outgoing boundary conditions. The steady-state flux through a selected continuum state $|j_0\rangle$, corresponding to a detector that resolves a given outgoing energy and direction, is then given by $\eta|d_{j_0}|^2$. Scaled by $|c_{in}|^2$, we define
\begin{equation}
    F_{out}\equiv\eta|d_{j_0}/c_{in}|^2.
\end{equation}
The resulting steady-state spectrum is
\begin{equation}
    F_{out}=\eta\left|\sum_k\frac{V_{k,j}}{E_{in}-E_{out}+i\eta/2}\frac{V_{in}}{(E_{in}-E_M)-\sum_{j,k}|V_{k,j}|^2[E_{in}-E_{j}+i\eta/2]^{-1}}\right|^2,
    \label{eq:Foutnocav}
\end{equation}
where $E_{out}\equiv E_{j_0}$, and for identical molecules we write $E_k=E_M$. A detailed derivation is given in Sec. I of the Supplementary Information (SI).

\subsection{Molecules inside a cavity}
When the molecular cluster is placed inside a cavity, we consider two possible roles of cavity mode $|c\rangle$.

In the first scenario, the cavity acts as an intermediary between matter and radiation. The Hamiltonian is 
\begin{align}
    \hat{H}_{C1}&=E_{in}|in\rangle\langle in|+E_c|c\rangle\langle c|+V_{c,in}|c\rangle\langle in|+V_{in,c}|in\rangle\langle c|\notag\\
    &+\sum_k\left[E_k|k\rangle\langle k|+V_{k,c}|k\rangle\langle c|+V_{c,k}|c\rangle\langle k|\right]+\sum_{j}\left[E_j|j\rangle\langle j|+V_{c,j}|c\rangle\langle j|+V_{j,c}|j\rangle\langle c|\right],
    \label{eq:HC1}
\end{align}
Here $V_{in,c}=V_{c,in}^*,V_{c,k}=V_{k,c}^*$. The corresponding scattering spectrum takes the form
\begin{equation}
    F_{out}=\frac{\eta|V_{c,j}|^2}{(E_{out}-E_{in})^2+(\eta/2)^2}\left|\frac{V_{in,c}}{(E_{in}-E_c)-\sum_k|V_{c,k}|^2[E_{in}-E_M]^{-1}-\sum_j|V_{c,j}|^2[E_{in}-E_j+i\eta/2]^{-1}}\right|^2.
    \label{eq:Foutcavity1}
\end{equation}
In the second scenario, the cavity couples only to the molecular states and not directly to the incident or outgoing radiation channels. The Hamiltonian is
\begin{align}
    \hat{H}_{C2}&=\hat{H}_{SS}+E_c|c\rangle\langle c|+\sum_k(V_{c,k}|c\rangle\langle k|+V_{k,c}|k\rangle\langle c|).
    \label{eq:HC2}
\end{align}
and the corresponding flux is
\begin{equation}
    F_{out}=\eta\left|\sum_k\frac{V_{k,j}}{E_{in}-E_{out}+i\eta/2}\frac{V_{in,k}}{(E_{in}-E_M)-\sum_{k'}|V_{c,k'}|^2[E_{in}-E_{c}]^{-1}-\sum_{j,k'}|V_{k^\prime,j}|^2[E_{in}-E_{j}+i\eta/2]^{-1}}\right|^2.
    \label{eq:Foutcavity2}
\end{equation}
Technique details of these derivations are given in Sec. I in SI.

\subsection{Stochastic approach}
The damping associated with the continuum primarily determines the width of the elastic peak, whereas broadening of the fluorescence peak is mainly induced by thermal fluctuations from the molecular environments. To describe this effect, we model each molecular excitation energy $E_k(t)$ as a stochastic process with mean value is $\langle E_k(t)\rangle=E_M$ and with exponentially decaying temporal correlations, $\langle E_k(t_1)E_{k'}(t_2)\rangle=\delta_{kk'}\sigma^2e^{-|t_1-t_2|/\tau_c}$, where $\delta_{kk'}$ is the Kronecker delta, $\tau_c$ is the correlation time and sigma sets the amplitude \cite{Rybicki1995,Gillespie1996,Chen2022}.

To generate such trajectories numerically, we discretize time as $\{t_j\},t_j=j\Delta t$ and define $x_j=x_j(t_j)$, where $x(t)$ denotes a stochastic variable such as the fluctuations of a molecular excitation energy. The joint probability distribution is written as
\begin{equation}
   \text{Prob}(x_1,...,x_n)=\text{Prob}(x_1)\prod_{i=2}^n\text{Prob}(x_i|x_{i-1}).
\end{equation}
The initial distribution is taken to be Gaussian,
\begin{equation}
    \text{Prob}(x_1)=\frac{1}{\sqrt{2\pi\sigma^2}}\exp\left[-\frac{x^2_1}{2\sigma^2}\right]
\end{equation}
with the conditional distribution
\begin{equation}
    \text{Prob}(x_i|x_{i-1})=\frac{1}{\sqrt{2\pi\sigma^2(1-r_c^2)}}\exp\left[-\frac{(x_i-r_cx_{i-1})^2}{2\sigma^2(1-r^2)}\right],
\end{equation}
with $r_c=\exp(-\Delta t/\tau_c)$. Thus, $x_i$ is Gaussian distributed with mean $\langle x_i\rangle=x_{i-1}e^{-\Delta t/\tau_c}$ and variance $\sigma^2(1-e^{-2\Delta t/\tau_c})$. In the slow modulation limit (large $\tau_c$), the fluctuations become nearly static over the timescale of the system dynamics.

To include thermal fluctuations in the calculated steady-state spectra, we evaluate the flux for each stochastic realization of the molecular onsite energies, then perform an ensemble average, followed by a time average over one driving period $T=2\pi\hbar/E_{in}$ in the long-time regime.

\subsection{Density-matrix approach}
As an alternative treatment of relaxation through the radiation continuum, we first eliminate the continuum amplitudes. From Eq. \eqref{eq:schSSF}, we obtain
\begin{equation}
    d_j(t)=-\frac{i}{\hbar}\sum_kV_{k,j}\int_0^t e^{-iE_j(t-\tau)/\hbar}c_k(\tau) d\tau.
    \label{eq:djt}
\end{equation}
where we have assumed the continuum is initially unpopulated, $d_j(0)=0$. Writing 
\begin{equation}
    c_k(t)=\tilde{c}_k(t)e^{-iE_Mt/\hbar}
\end{equation}
and substituting Eq. \eqref{eq:djt} into the equation of $c_k(t)$, we obtain
\begin{equation}
    \frac{d\tilde{c}_k}{dt}=-\frac{i}{\hbar}V_{in,k}c_{in}e^{iE_M/\hbar}-\frac{i}{\hbar}V_{k,j_0}d_{j_0}e^{iE_Mt/\hbar}-\frac{1}{\hbar}\sum_{k'}\tilde{c}_{k'}(t)\left(i\Lambda(E_M)+\frac{1}{2}\Gamma(E_M)\right),
    \label{eq:selfshift}
\end{equation}
with the self-energy terms
\begin{align}
    \label{eq:Lambda}
    \Lambda(E)&\equiv\mathcal{P}\int\frac{\overline{|V_{k,j}|^2\rho(E_j)}}{E-E_j} dE_j,\\
    \Gamma(E)&\equiv2\pi(\overline{|V_{k,j}|^2\rho(E_j))}|_{E_j=E}.
    \label{eq:Gamma}
\end{align}
where $\mathcal{P}$ denotes the principal value and $\rho(E)$ is the density of states in the continuum. The coupling of the molecular states to the common radiative continuum therefore generates both energy shift $\Lambda(E)$ and a radiative decay rate $\Gamma(E)$. Using the fact that the continuum is dense, we select one state $|j_0\rangle$ from $\{j\}$ and identify it as the detected outgoing channel, denoted $|out\rangle$, with energy $E_{out}\equiv E_{j_0}$. Following Ref. \cite{Nitzan}, we assume that 
\begin{equation}
    \sum_{j\neq j_0} e^{iE_j(t-\tau)/\hbar}V_{k,j}V_{k',j}
\end{equation}
is sharply peaked at $t = \tau$ so that $c_{k'}(\tau)$  may be taken outside the integral and the lower limit can be extended to $t\rightarrow-\infty$. Transforming back to $c_k$, we obtain
\begin{equation}
    \frac{dc_k}{dt}=-\frac{iE_M}{\hbar}c_k-\frac{i}{\hbar}V_{in,k}c_{in}-\frac{i}{\hbar}V_{k,out}d_{out}-\frac{i}{\hbar}\sum_{k'}c_{k'}\left(\Lambda(E_M)-\frac{i}{2}\Gamma(E_M)\right).
\end{equation}

To incorporate thermal effect, we work in Liouville space and construct the density matrix $\hat{\rho}$ with elements $\rho_{\alpha,\beta}(t)=c_{\alpha}(t)c^*_{\beta}(t),\alpha,\beta=in,out,k$. Its evolution is taken to obey
\begin{align}
    i\hbar\frac{d\hat{\rho}}{dt}&=[\hat{H},\hat{\rho}]-\frac{i}{2}(\hat{\rho}\mathbf{\Gamma}+\mathbf{\Gamma}\hat{\rho})-i\gamma*\hat{\rho},
    \label{eq:liuvilledamp}
\end{align}
where the small energy shift $\Lambda$ is neglected. The damping matrix $\mathbf{\Gamma}$ has dimension $(N+2)\times(N+2)$; its entries $(k,k')$ are taken equal to $\Gamma$, while all other entries involving incident and outgoing channels are zero \footnote{We also make the wide band approximation by assuming that the edge of the continuum is far from the energetic region of interest, through which the density of states is constant.}. Thermal dephasing is introduced phenomenologically through the operator $\gamma*$, which acts as follows: Each $\rho_{\alpha,k}$ and $\rho_{k,\alpha}$ with $\alpha=in,out$ acquires a term $-i\gamma/2$, and each intermolecular coherence $\rho_{k,k'},k'\neq k$ acquires a term $-i\gamma$ \footnote{Diagonal elements $\rho_{k,k}$ do not receive the relaxation rate $\gamma$ because the dephasing is associated with transitions in molecular energy spacing, while $\rho_{k,k}$ refers to the same molecular state.}. To impose outgoing boundary conditions on $|out\rangle$, we additionally assign small damping rates $-i\eta/2,-i\eta/2$ and $-i\eta$ to $\rho_{in,out},\rho_{out,in}$ and $\rho_{out,out}$, respectively. Explicit equations for the case $N=2$ are given in Sec. II in SI.

When the molecular cluster is placed inside the cavity, we again consider two cavity roles defined by Eqs. (\ref{eq:HC1}) and (\ref{eq:HC2}). Relative to the no-cavity case, the Hilbert space is enlarged by one cavity state, so the corresponding matrices have dimensions $(N+3)\times(N+3)$. For the Hamiltonian $\hat{H}_{C2}$, the Liouville equation retains the same structure as Eq. \eqref{eq:liuvilledamp}, except that the dephasing operator also acts on $\rho_{c,k}$ and $\rho_{k,c}$. For $\hat{H}_{C1}$, because the cavity mode alone mediates the coupling between the incident and outgoing radiation channels, the damping matrix has only one nonzero element, corresponding to the cavity-state population.

Finally, to compute the steady-state flux, we solve 
\begin{equation}
    d\hat{\rho}/dt=0
\end{equation}
 while keeping $\rho_{in,in}$ fixed, which is equivalent to the driving boundary condition $c_{in}(t)=\tilde{c}_{in}\exp(-iE_{in}t/\hbar)$. The scaled outgoing flux is then evaluated as 
 \begin{equation}
      F_{out}=\eta\left|\frac{\rho_{out,out}}{\rho_{in,in}}\right|.
 \end{equation} 

\section{Results}
\begin{figure}[!htp]
\includegraphics[width=0.8\textwidth]{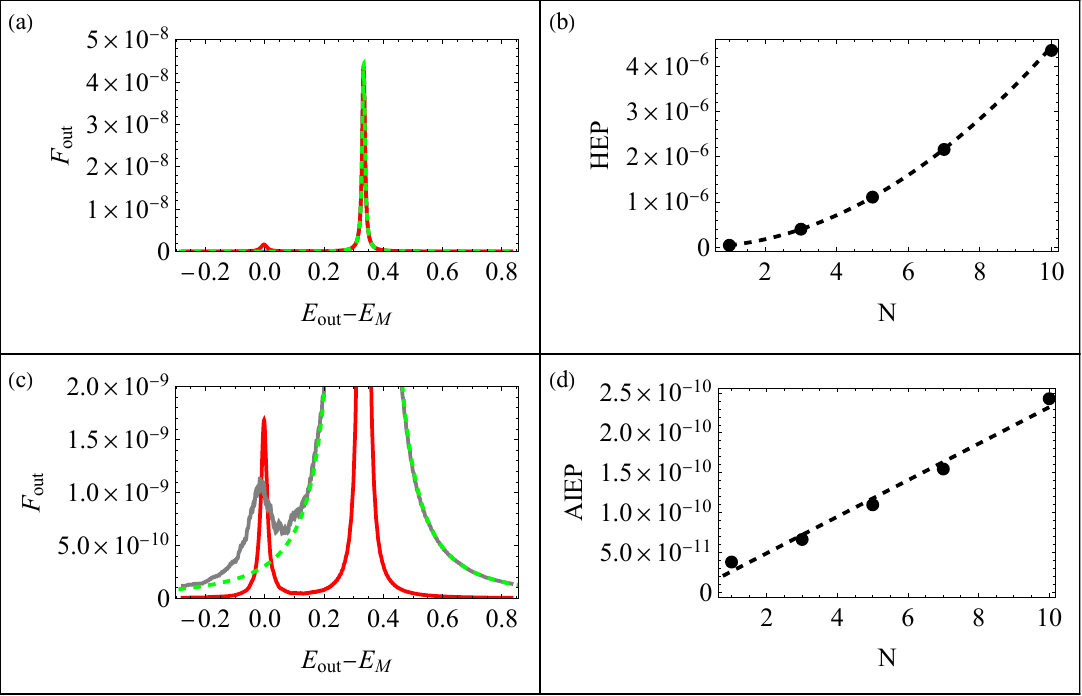}
\caption{Long-time scattering spectra of a molecular ensemble outside the cavity under continuous wave (CW) pumping at $E_{in}=4/3$. Panel (a) depicts the spectrum for a single molecule (red solid line), together with the analytic fit from Eq. \eqref{eq:Foutnocav} (green dashed line). Panel (b) records the height of the elastic peaks (HEP) as a function of the number of molecules $N$; the dashed line is a quadratic fit. Panel (c) shows inelastic fluorescence spectra for $N=1$ (red solid line) and $N=5$ (gray solid line); the $N=5$ result is compared with Eq. \eqref{eq:Foutnocav} (green dashed line). Panel (d) indicates the area of the inelastic emission peak (AIEP) extracted from panel (c) as a function of $N$; the dashed line is a linear fit. The spectra are averaged over 200 stochastic realizations with the correlation time $\tau_c=9/200$, and onsite energy fluctuation amplitude $\sigma=1/9$. All coupling strengths are $1/900$, and the damping parameter is $\eta=1/90$. Energies are given in units of $E_M$.}
\label{fig:stocout}
\end{figure}

\begin{figure}[!htp]
\includegraphics[width=0.8\textwidth]{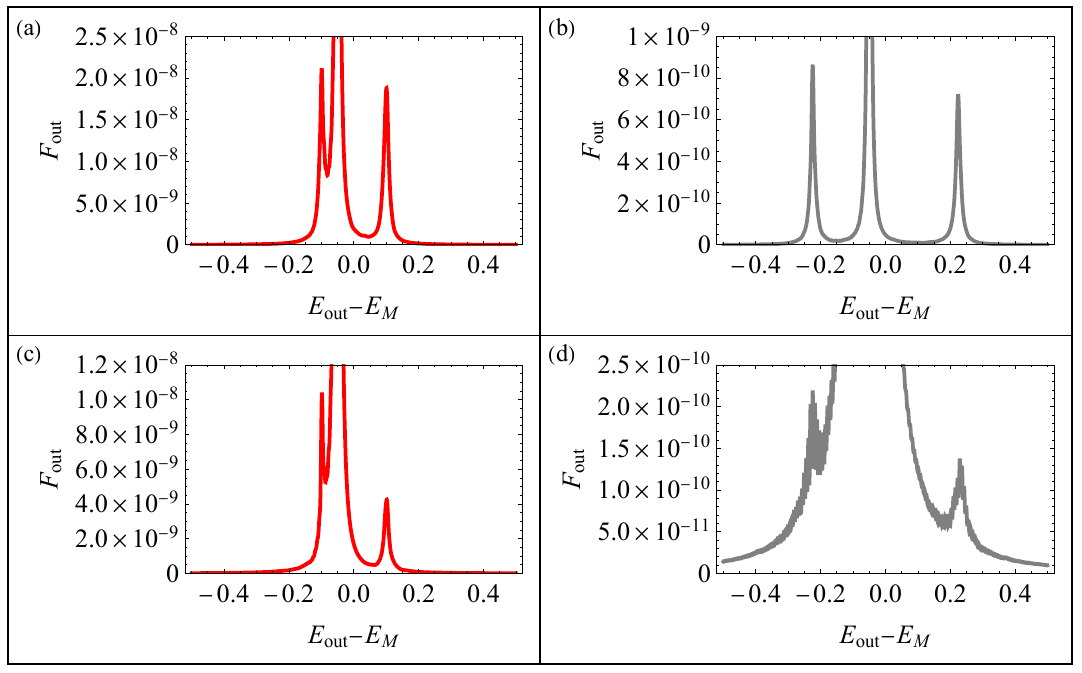}
\caption{Long-time scattering spectra of a molecular ensemble inside a cavity under CW pumping at $E_{in}=4/3$. Panels (a) and (b) show spectra for $N=1$ (red line) and $N=5$ (gray line), respectively, for the model in which the cavity mode couples to both molecules and the radiation channels [Eq. \eqref{eq:HC1}]. Panels (c) and (d) present spectra for $N=1$ (red line) and $N=5$ molecules (gray line), respectively, for the model in which the cavity mode couples only to the molecules [Eq. \eqref{eq:HC2}]. The spectra are averaged over 200 stochastic realizations with the correlation time $\tau_c=1/20$, and onsite energy fluctuation amplitude $\sigma=1/10$. The cavity is in resonance with the molecule transition, $E_c=1$. The cavity-molecule coupling strength is $0.1$, all other coupling parameters are $0.001$ and the damping parameter is $\eta=0.01$. Energies are given in units of $E_M$.}
\label{fig:stocin}
\end{figure}

\begin{figure}[!htp]
\centering
\includegraphics[width=0.8\textwidth]{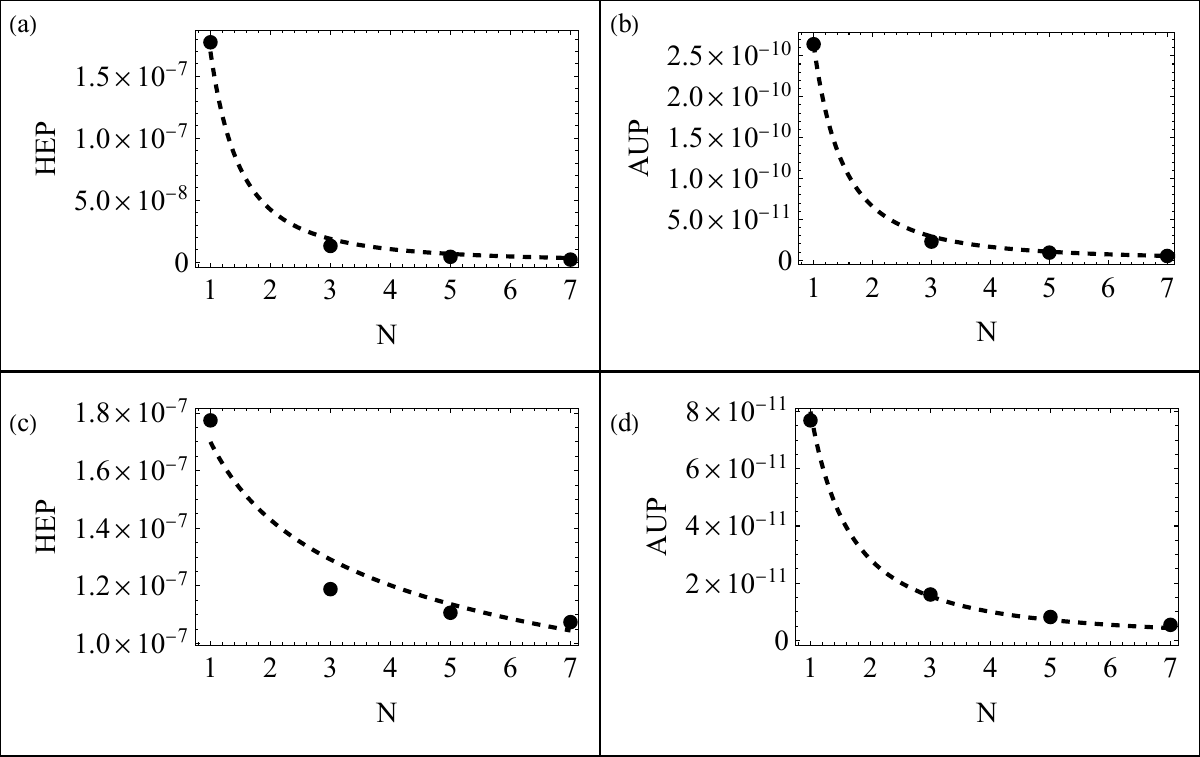}
\caption{Collective scaling of elastic and polaritonic spectral features for molecules inside the cavity, obtained from the stochastic approach. Panels (a) and (b) show the HEP and the area of the upper-polariton peak (AUP), respectively, as functions of the number of molecules $N$, for the model in which the cavity bridges molecules and the radiation continuum. The dashed lines are fits proportional to $N^{-2}$. Panels (c) and (d) depict the HEP and AUP, respectively, as functions of $N$, for the model in which the cavity couples only to the molecules. The dashed lines are fits proportional to $\sim N^{-0.25}$ in panel (c) and $\sim N^{-1.5}$ in panel (d). All parameters are same as Fig. \ref{fig:stocin}.}
\label{fig:stoAIEP}
\end{figure}

In this section, we present the scattering spectra obtained using two approaches introduced above. Unless otherwise specified, the cavity mode is taken to be resonant with the molecular excitation energy $E_c=E_M$, and we set $\hbar=1$. All energies are reported in units of $E_M$.

Steady-state fluxes calculated using the stochastic approach are shown in Fig. \ref{fig:stocout} for molecules outside the cavity and Fig. \ref{fig:stocin} for molecules inside the cavity. Figure. \ref{fig:stocout}(a) displays the spectrum of a single molecule driven at $E_{in}=4/3$. Two distinct features are observed: an elastic Rayleigh peak at $E_{out}=E_{in}$ and an inelastic fluorescence peak near $E_{out}=E_M$. The elastic peak is well-fitted by the analytical expression in Eq. \eqref{eq:Foutnocav}. As shown in Fig. \ref{fig:stocout}(b), its height scales approximately as $\sim N^2$, with the number of molecules, indicating a collective coherent contribution to the elastic response. The inelastic fluorescence feature is illustrated in Fig. \ref{fig:stocout}(c), where spectra for $N=1$ and $N=5$ are compared. In the regime with relatively fast thermal modulation, the width and height of this feature are controlled mainly by the combination $\sigma^2\tau_c$ (see Fig. S1(a) in SI). To quantify the inelastic response, we define the area of the inelastic emission peak (AIEP) by integrating the spectral weight in the fluorescence region away from the dominant elastic peak. Figure \ref{fig:stocout}(d) shows that the AIEP increases approximately linearly with $N$, consistent with an incoherent contribution from individual molecular excitations \footnote{The elastic peak is dominant in the overall shape of the lineshape, it may cause distortion to the closer wing or even suppress the inelastic peak when the number of molecules is large (see Fig. S1(b) in SI).}

Figure \ref{fig:stocin} shows the corresponding stochastic spectra when the molecules are placed inside the cavity. We consider separately the two cavity configurations introduced in Sec. II: in Figs. \ref{fig:stocin}(a) and \ref{fig:stocin}(b), the cavity mode couples both to the molecular states and to the radiation channels [Eq. (\ref{eq:HC1})], whereas in Figs. \ref{fig:stocin}(c) and \ref{fig:stocin}(d), it couples only to the molecular states [Eq. (\ref{eq:HC2})]. In both cases, the fluorescence feature that appears near $E_M$ outside the cavity is replaced by two peaks, corresponding to the upper and lower polaritons. The elastic components of the spectra are well reproduced by the analytical expressions in Eqs. (\ref{eq:Foutcavity1}) and (\ref{eq:Foutcavity2}); the full comparison is shown in Fig. S2 in SI.  The splitting between the upper and lower polaritonic peaks reflects the collective light-matter coupling and increases with the number of molecules in a manner consistent with the usual $\sqrt{N}$ Rabi splitting. To characterize how the spectral weight is redistributed with system size, Fig. \ref{fig:stoAIEP} shows the height of the elastic peak (HEP) and the area of the upper-polariton peak (AUP) as functions of $N$. For the configuration in which the cavity mediates the coupling between molecules and radiation, both the HEP and AUP decrease with increasing $N$, with the fitted trends shown in Figs. \ref{fig:stoAIEP}(a) and \ref{fig:stoAIEP}(b). When the cavity couples only to the molecules, the suppression with increasing system size is weaker for the elastic peak [Fig. \ref{fig:stoAIEP}(c)] but remains pronounced for the upper-polaritonic weight [Fig. \ref{fig:stoAIEP}(d)]. Similar trends are found for the lower-polariton area (ALP); see Fig. S3 in SI.

\begin{figure}[H]
\centering
\subfloat[][]{
\begin{minipage}[t]{0.8\textwidth}
\flushleft
\includegraphics[width=0.8\textwidth]{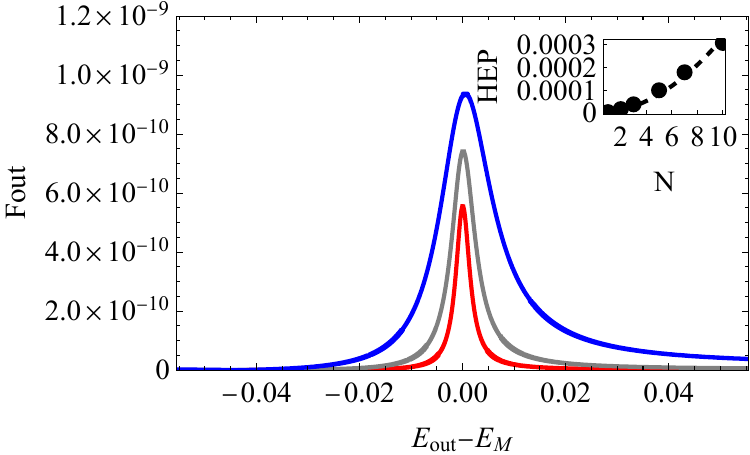}
\end{minipage}
}
\\
\subfloat[][]{
\begin{minipage}[t]{0.8\textwidth}
\flushleft
\includegraphics[width=0.8\textwidth]{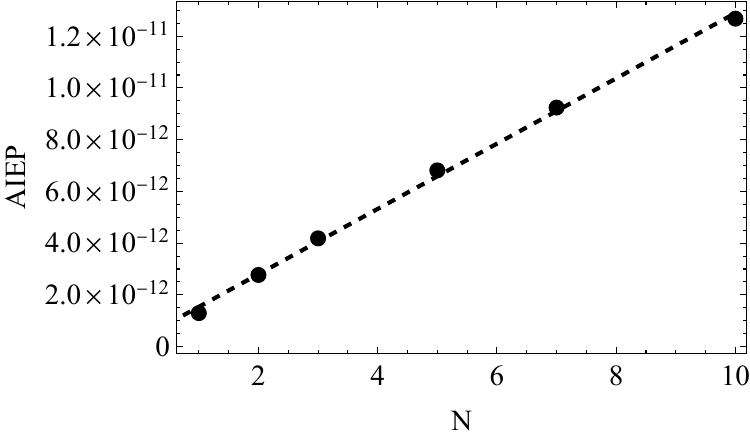}
\end{minipage}
}
\caption{Scattering spectra obtained from the steady-state density-matrix approach for molecules outside the cavity. Panel (a) displays the energy-resolved outgoing flux under CW pumping at $E_{in}=4/3$. Red, gray and blue lines correspond to $N=1,2,5$ molecules, respectively. The inset shows the HEP as a function of $N$; the dashed line is a quadratic fit. Panel (b) shows AIEP as a function of $N$; the dashed line is a linear fit. The continuum-induced damping is $\Gamma=1/9000$, which is also taken as the damping $\eta$ of the outgoing state, and the dephasing rate is $\gamma=0.1\Gamma$. All coupling strengths are $1/900$. Energies are given in units of $E_M$.}
\label{fig:DMout}
\end{figure}

\begin{figure}[!htp]
\centering
\subfloat[][]{
\begin{minipage}[t]{0.68\textwidth}
\flushleft
\includegraphics[width=0.68\textwidth]{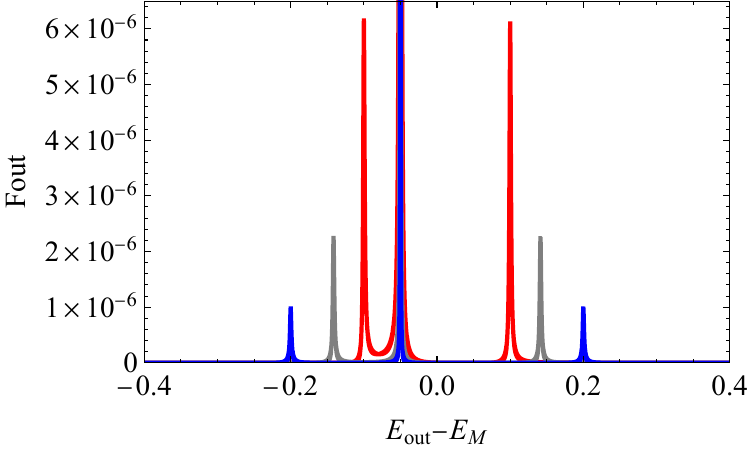}
\end{minipage}
}
\\
\subfloat[][]{
\begin{minipage}[t]{0.68\textwidth}
\flushleft
\includegraphics[width=0.68\textwidth]{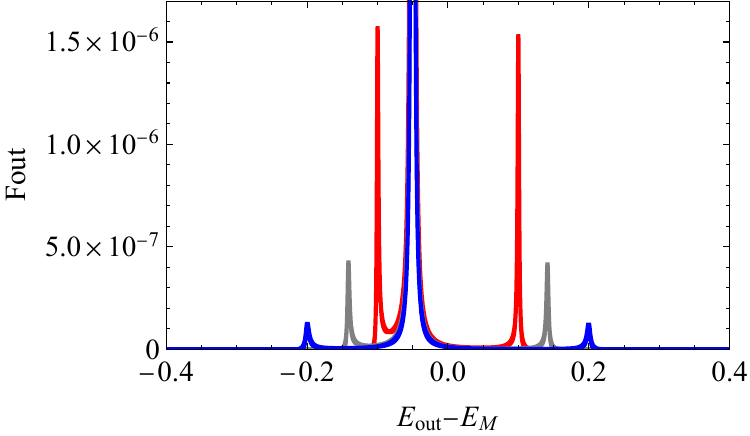}
\end{minipage}
}
\caption{Scattering spectra obtained from the steady-state density-matrix approach for molecules inside the cavity. Panel (a) displays the outgoing energy flux under CW pumping at $E_{in}=0.95$ for the model in which the cavity mode couples to both the molecules and the radiation channels [Eq. \eqref{eq:HC1}]. Panel (b) presents the corresponding result for the model in which the cavity mode couples only to the molecules [Eq. \eqref{eq:HC2}]. The continuum-induced damping is $\Gamma=0.0001$, which is also taken as the damping $\eta$ of the outgoing state, and the dephasing rate is $\gamma=0.1\Gamma$. The cavity is in resonance with the molecule transition, $E_c=1$. The cavity-molecule coupling is $0.1$, and all other couplings are $0.001$. Energies are given in units of $E_M$.}
\label{fig:DMin}
\end{figure}

\begin{figure}[!htp]
\includegraphics[width=0.8\textwidth]{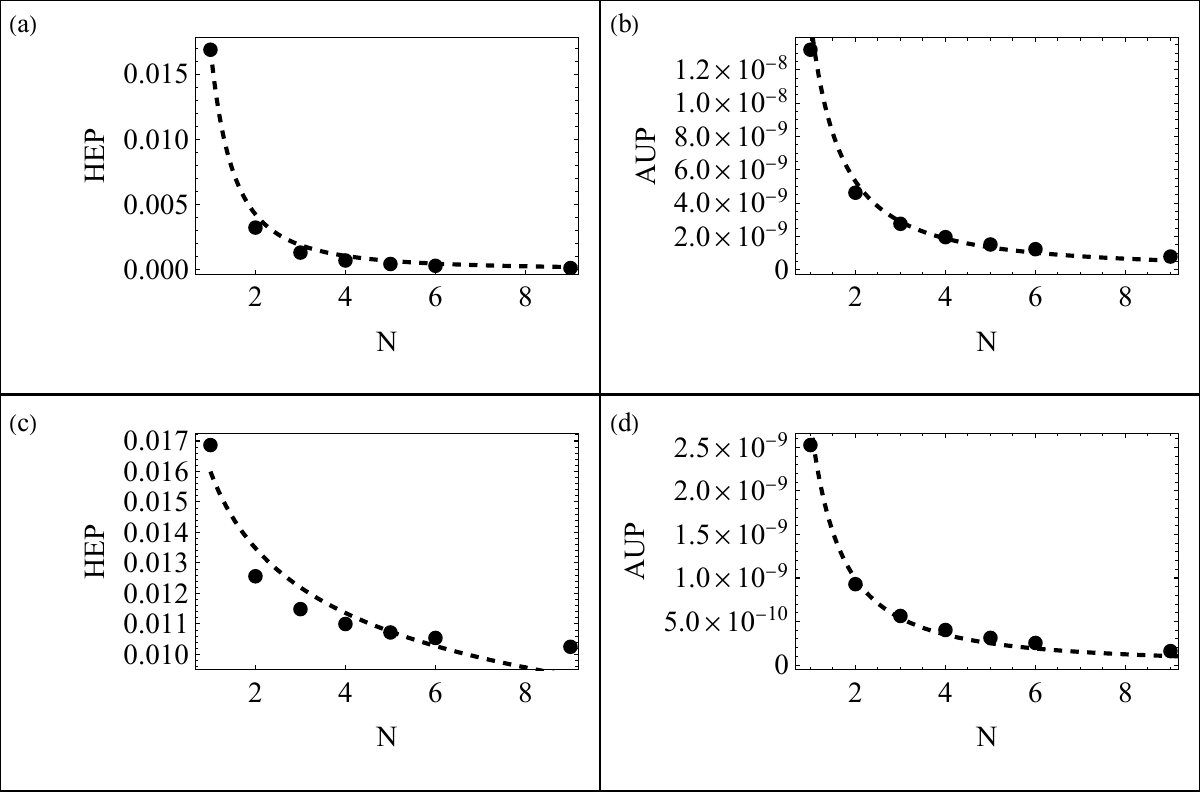}
\caption{Collective scaling of elastic and polaritonic spectral features for molecules inside the cavity, obtained from the density-matrix approach. Panels (a) and (b) show HEP and AUP, respectively, as functions of the number of molecules $N$, for the model in which the cavity couples to both the molecules and the radiation continuum. The dashed lines are fits proportional to $N^{-2}$ in panel (a) and $ N^{-1.5}$ in panel (b). Panels (c) and (d) depict the HEP and AUP, respectively, as functions of $N$, for the model in which the cavity mode couples only to the molecules. The dashed lines are fits proportional to $N^{-0.25}$ in panel (c) and $ N^{-1.5}$ in panel (d). All parameters are the same as in Fig. \ref{fig:DMin}.}
\label{fig:DMAIEP}
\end{figure}

We next turn to the density-matrix approach. Figure \ref{fig:DMout} shows the spectra obtained from the steady-state solution of the Liouville equation for molecules outside the cavity. As in the stochastic calculations, the elastic peak dominates the overall lineshape, so in Fig. \ref{fig:DMout}(a) the full spectrum is shown together with an inset highlighting the HEP as a function of $N$. The elastic peak again exhibits an approximately quadratic scaling with the number of molecules. Figure \ref{fig:DMout}(b) shows that the AIEP grows approximately linearly with $N$, in agreement with the behavior found in the stochastic treatment. Additional spectra for different pumping conditions are presented in Fig. S4 in SI.

Figure \ref{fig:DMin} presents the spectra obtained from the density-matrix approach for molecules inside the cavity. Panel (a) corresponds to the model in which the cavity couples to both molecules and radiation channels, whereas panel (b) corresponds to the model in which the cavity couples only to the molecules. In both cases, the inelastic spectral feature splits into upper- and lower-polaritonic peaks, consistent with the stochastic results. The corresponding system-size dependence of the HEP and AUP is summarized in Fig. \ref{fig:DMAIEP}. The behavior of the lower-polaritonic area is shown in Fig. S5 in SI.

Overall, the two approaches yield qualitatively similar physical trends. Outside the cavity, the elastic response scales quadratically with $N^2$, whereas the integrated inelastic fluorescence weight grows roughly linearly with $N$. Inside the cavity, the inelastic peak is replaced by polaritonic doublets, and both the elastic and polaritonic spectral weights display nontrivial dependence on molecular number that depends on the role played by the cavity mode. At the same time, the two methods differ in the detailed shape and relative intensity of the polaritonic peaks, an issue that we analyze further in the next section.

\section{Discussion}
Outside the cavity, the approximately quadratic growth of the elastic peak height with the number of molecules is reminiscent of superradiant scaling. The behavior reflects the coherent addition of molecular dipole amplitudes in the bright collective state. For identical molecules, the collective transition dipole associated with elastic scattering scales as \cite{Nitzan}
\begin{equation}
    |\langle B|\hat{\mu}|B\rangle|^2\propto N^2,
\end{equation}
where $|B\rangle=\sum_k|k\rangle/\sqrt{N}$ is the molecular bright state (single exciton state) and $\hat{\mu}=\sum_k\hat{\mu}_k$ is the total dipole operator. In this sense, the elastic Rayleigh response retains the phase coherence of the incident field and therefore exhibits a collective enhancement with increasing system size. The inelastic fluorescence feature behaves differently. In both the stochastic and density-matrix treatments, its integrated spectral weight grows approximately linearly with $N$, indicating that it is dominated by local relaxation processes acting on individual molecular excitations. Because the molecules are taken to be noninteracting, thermal relaxation and dephasing act independently on each molecule and do not generate the same coherent enhancement as in the elastic channel. The distinction between the $N^2$ scaling of the elastic peak and the nearly linear scaling of the fluorescence weight therefore provides a simple way to separate coherent collective response from incoherent thermally assisted emission in this model. 

The two approaches clarify this point from different perspectives. In the stochastic treatment, the scattering probability in \eqref{eq:Foutnocav}, is built from coherent amplitudes associated with excitation and re-emission through the molecular manifold. In the absence of thermal fluctuations, this framework captures the elastic response but does not, by itself, generate a fluorescence peak at the molecular excitation energy. The appearance of such an inelastic feature, therefore, reflects the stochastic modulation of the molecular transition energies by the thermal environment. In the density-matrix treatment, by contrast, the common radiative continuum induces collective decay through the damping matrix $\Gamma$, while thermal dephasing is introduced phenomenologically through the parameter $\gamma$. The elastic and inelastic features are then shaped jointly by collective radiative coupling and local dephasing.

When the molecular ensemble is placed inside the cavity, the optical response is modified by hybridization between molecular excitations and the cavity mode. At zero detuning, $E_c=E_M$, the inelastic feature near the molecular transition splits into upper- and lower-polaritonic peaks, as expected in the strong-coupling regime. Under parameter choices considered here and for fixed cavity-molecule coupling, the heights or integrated weights of these peaks decrease with increasing $N$. The elastic peak shows a similar suppression. Physically, the cavity introduces additional pathways through which the excitation is redistributed among collective light-matter modes and outgoing channels. The resulting spectra weight observed in any one selected output channel is therefore reduced as the system size increases. This reduction is more pronounced when the cavity also couples directly to the radiation continuum, because in that case, the cavity plays a more active role in both light-matter hybridization and radiative out-coupling.

Although the stochastic and density-matrix approaches agree qualitatively, some quantitative differences remain, especially in the polaritonic region when the cavity couples both to the molecular ensemble and to the continuum. For example, the scaling of the polaritonic peak areas obtained from the two methods is not identical. A likely reason is that, in the density-matrix treatment the dissipation induced by the continuum is represented by an effective damping term, which compresses the influence of many radiative channels into a simplified Markovian description.  This approximation may underestimate part of the collective structure captured more explicitly in the stochastic wavefunction calculations.

Finally, we emphasize that the collective trends reported here for cavity-modified spectra apply to the case of fixed cavity-molecular coupling. Different conclusions may be reached if one instead fixes the collective Rabi splitting while varying the number of molecules, as illustrated in Figs. S6 and S7 in SI. The interpretation of collectivity in cavity-modified scattering spectra must therefore be tied to the specific scaling protocol and model assumptions used.

\section{Conclusion}
In this work, we have studied steady-state optical scattering from ensembles of noninteracting molecules, both outside and inside an optical cavity, while incorporating local thermal relaxation. We employed two complementary theoretical approaches: a stochastic description based on time-dependent fluctuations of the molecular excitation energies, and a density-matrix description based on phenomenological relaxation and dephasing rates. Although these two methods differ in their microscopic assumptions, they lead to qualitatively consistent physical trends.

Outside the cavity, the spectra exhibit an elastic Rayleigh peak at the incident-photon energy and an inelastic fluorescence feature near the molecular excitation energy. The elastic peak height scales approximately quadratically with the number of molecules, reflecting coherent collective enhancement, whereas the integrated fluorescence signal scales approximately linearly, consistent with incoherent local thermal relaxation acting on individual molecules.

Inside the cavity, the inelastic feature near the molecular excitation energy splits into upper- and lower-polaritonic peaks in the strong-coupling regime. Under fixed cavity-molecule coupling, both the elastic and polaritonic spectral weights show characteristic dependence on the number of molecules, with stronger suppression observed when the cavity also couples directly to the radiation continuum. These results illustrate how cavity hybridization and thermal relaxation jointly determine the observable scattering spectrum.

The comparison between the two approaches also clarifies their respective strengths and limitations. The stochastic treatment naturally captures thermally induced fluctuations and reproduces detailed-balance-like asymmetry between upper and lower polaritons, but for large systems the fluorescence feature can be strongly obscured by the dominant elastic peak. The density-matrix treatment is computationally more convenient for larger systems and reproduces the main collective trends, but its phenomenological form does not fully encode detailed balance and may smooth out part of the collective structure associated with the continuum.

Overall, our results help distinguish collective coherent contributions from local incoherent relaxation in cavity-modified scattering spectra. At the same time, the present model remains intentionally minimal. In particular, vibronic structure is not included, so the calculated inelastic spectra should be interpreted as fluorescence-like emission rather than resonance Raman scattering. Extensions that include vibrational degrees of freedom, excitation transport, disorder-induced localization, and more microscopic treatments of dissipation should provide a broader framework for analyzing collective spectroscopy in molecular cavities.

\section*{Acknowledgements}
Discussions with Abraham Nitzan and Michael Galperin are greatly acknowledged. This work was financially supported by the National Natural Science Foundation of China (No. 12404232), start-up funding from the Chinese University of Hong Kong, Shenzhen (No. UDF01003468) and the Shenzhen City “Pengcheng Peacock” Talent Program.

\section*{Statements and Declarations}

\subsection*{Conflict of interest}
The author declares no conflict of interest.

\section*{Data availability}
The data that support the findings of this study are available from the corresponding author upon reasonable request.

\bibliography{reference}

\end{document}